\documentclass[
 reprint,
twocolumn,
 amsmath,amssymb,
 aps,
]{revtex4}

\usepackage{graphicx}
\usepackage{dcolumn}
\usepackage{bm}
\usepackage[normalem]{ulem}
\usepackage{soul}
\usepackage{color, xcolor}

\begin{document}

\preprint{APS/123-QED}

\title{Coarse-graining conformational dynamics with multi-dimensional generalized Langevin equation: how, when, and why}

\author{Pinchen Xie}
 \affiliation{Program in Applied and Computational Mathematics, Princeton University, Princeton, NJ 08544, USA}

\author{Yunrui Qiu}
\affiliation{
 Department of Chemistry, University of Wisconsin-Madison, Madison, WI 53706, USA
}%

\author{Weinan E}%
\affiliation{%
 AI for Science Institute, Beijing 100080, China}%
\affiliation{Center for Machine Learning Research and School of Mathematical Sciences, Peking University, Beijing 100084, China}

\date{\today}

\begin{abstract}
A data-driven ab initio generalized Langevin equation (AIGLE) approach is developed to learn and simulate high-dimensional, heterogeneous, coarse-grained conformational dynamics. Constrained by the fluctuation-dissipation theorem, the approach can build coarse-grained models in dynamical consistency with all-atom molecular dynamics. We also propose practical criteria for AIGLE to enforce long-term dynamical consistency. Case studies of a toy polymer, with 20 coarse-grained sites, and the alanine dipeptide, with two dihedral angles, elucidate why one should adopt AIGLE or its Markovian limit for modeling coarse-grained conformational dynamics in practice.

\end{abstract}


\maketitle

\section{Introduction} 

Modeling changes in molecular conformation is crucial across various scientific domains. Unbiased all-atom molecular dynamics (MD) is a potent tool for simulating conformational dynamics with atomistic precision and femtosecond resolution. However, the dynamic processes of interest often unfold over the timescale of milliseconds or longer, far surpassing the timescale that MD simulations can accommodate~\cite{kubelka2004protein}. 
To overcome the limitation of timescale, one may instead simulate a coarse-grained (CG) model derived from an all-atom model. The generalized coordinates, $X$, of the CG model are identified with the CG variables $f_X(q)$ of the all-atom model. $f_X(q)$ is a vector-valued function of atomic coordinates $q$. 
 
The consistency between a CG model and an all-atom model requires efforts at two levels: thermodynamic consistency (TC) and dynamical consistency (DC). Let $H(q)$ be the all-atom potential energy surface and $G(X)$ be the effective Hamiltonian, or free energy, of the CG system. TC is specified by $e^{-\beta G(X)} \propto \int e^{-\beta H(q)}\delta(f_X(q)-X) dq$. Force matching~\cite{izvekov2005multiscale} and metadynamics~\cite{barducci2008well} are typical schemes for extracting $G(X)$ from $H(q)$. DC is prescribed by the Mori-Zwanzig formalism~\cite{zwanzig2001nonequilibrium}, which specifies an effective equation of motion of $X$ that can recover the dynamical behavior of the CG variables in the atomic system. In practice, the exact equation of motion is intractable. Simple surrogate models are used as ansatz. Typical surrogate models incorporate TC and assume the form of $M\ddot{X}(t)=-\nabla_X G(X(t)) + \eta(t)$. $\eta$ is a non-conservative force representing a thermal bath. These models can be divided into two classes by whether $\eta$ is Markovian or non-Markovian. Typical examples are the Langevin equation (LE) and the generalized Langevin equation (GLE). Markovian LE splits $\eta(t)$ into a friction term $-\Lambda \dot{X}$ and a white noise term $w(t)$. GLE splits $\eta(t)$ into a memory term $\int_0^t K(s)\dot{X}(t-s)ds$ and a non-Markovian noise term $R(t)$. Both GLE and LE maintain the isothermal condition when constrained by the second fluctuation-dissipation theorem (2FDT)~\cite{kubo1966fluctuation}. LE, as a special case of GLE, is typically used for semi-quantitative characterization of CG dynamics. GLE is expected to have better quantitative accuracy since GLE can describe the exact CG dynamics of harmonic atomic systems~\cite{tuckerman2023statistical}. However, there is no consensus about how far GLE can enforce DC for general, anharmonic systems. This doubt, together with the difficulty of extracting GLE from high-dimensional all-atom MD data, and the lack of user-friendly software, discourage GLE from becoming a practical tool for modeling realistic systems. This paper strives to address some of these concerns. We want to answer three practical questions in the context of conformational dynamics: ``how'' to accurately parameterize multi-dimensional GLE from all-atom MD data, ``when'' can GLE enforce DC,  and  ``why'' should one prefer GLE over LE or reversely.

The question of why GLE should be used has been discussed for protein folding. Refs.~\cite{ayaz2021non, dalton2023fast} find GLE is capable of modeling with DC the CG model of one-dimensional (1D) CG variable associated with protein folding and unfolding, while LE is not able to reproduce simultaneously the folding and unfolding rate predicted by MD. These findings call for extensive studies of non-Markovian effects in general, multi-dimensional, CG models of biomolecules. Such investigation needs a set of data-driven algorithms, and software, for parameterization and simulation of GLEs.

Data-driven parameterization of GLE has been studied frequently over the past several decades~\cite{ MCCOY1975431, Berkowitz1981, Berkowitz1983, adelman1983chemical,  harris1990I, harris1990II, Horenko2007, fricks2009time, ceriotti2009langevin, ceriotti2009nuclear, ceriotti2010colored, ceriotti2010efficient, davtyan2015dynamic, lei2016data, li2017computing, jung2018generalized, lee2019, Santos2021, russo2022machine, lyu2023prl, xie2024ab}. However, most efforts are dedicated to low-dimensional cases, including the cases of identical 1D GLEs independently coupled to each degree of freedom of extensive homogeneous CG variables, such as identical CG particles in simple fluids~\cite{Berkowitz1983}. Modeling general CG models of large molecules poses greater challenges in two aspects. The first is simultaneous heterogeneity and high-dimensionality, default to modeling CG particles of biomolecules~\cite{kmiecik2016coarse}. The second is the coexistence of the short-term oscillatory behavior and the long-tail decaying behavior of the memory kernel $K(s)$. Note that the long-tail memory may exceed by far the timescale of local vibrational modes. 
Recent works on data-driven high-dimensional GLEs~\cite{lyu2023prl,xie2024ab} have started to address some of these difficulties by parameterizing memory kernels or noise generators with neural networks. While enhancing the accuracy of GLEs, the application of neural networks may reduce robustness and numerical efficiency for long-term simulation, where a balance of accuracy, efficiency, robustness, and practicality should be maintained. 

To address these concerns, we develop a practical numerical approach for data-driven GLE within the framework of extended Markovian dynamics~\cite{ceriotti2010colored}. Several approximations and constructions are made to deal with high dimensionality, heterogeneity, and short/long-term accuracy. Within this paper, we call the GLE extracted from all-atom MD data ab initio generalized Langevin equation (AIGLE). Furthermore, the LE derived from AIGLE is called an ab initio Langevin equation (AILE), to be distinguished from arbitrary ones.  A software package, AIGLETools~\cite{aigletools}, is developed to streamline the parameterization and the simulation of AIGLE/AILE for practical applications. The AIGLE/AILE approach has been explored previously in the context of lattice dynamics for solid-state materials~\cite{xie2024ab}. The approach developed here shares the same philosophy as the previous work but not the same methodological details, since conformational dynamics are heterogeneous.

In this work, we will discuss the "when" question by suggesting practical rules of thumb and explaining the possible failure of AIGLE due to the inappropriate choice of CG variables.
At last, we report case studies to elucidate the ``why'' questions in two classes of CG models. 
The first is particle-based coarse-graining, where adjacent atoms are combined into united particles that preserve the molecular topology~\cite{kmiecik2016coarse}. For this class, we choose a harmonic polymer model for demonstration.
The other class is collective variable-based coarse-graining.  Collective variables (CVs) are low-dimensional CG variables for characterizing specific reactions or conformational changes. Examples include dihedral angles and the root mean squared displacement (RMSD) associated with a reference structure. For this class, we study the alanine dipeptide molecule.
 
\section{Results}
\subsection{Construction of AIGLE and AILE}\label{sec:how}
We base AIGLE on orthogonally transformed generalized coordinates $Y=UX$. $X=X(t)$ is treated as a column vector of length $n$. The matrix $U$ diagonalizes the instantaneous velocity correlation matrix $\langle \dot{X}(t)\dot{X}^T(t)\rangle$. So $\langle \dot{Y}(t)\dot{Y}^T(t)\rangle$ is diagonal. $U$ eliminates instantaneous velocity cross-correlation and allows us to describe approximately the dynamics of each component of $Y$ with separated 1D GLEs, given as
\begin{equation}\label{gle}
\small
    \dot{P_i}(t) = F_i(t) + \int_0^t K_i(s)P_i(t-s) ds  + \sqrt{m_i}R_i(t).
\end{equation}
Here, $P_i(t)=m_i\dot{Y}_i(t)$ is the generalized momentum. $F_i(t)=-\partial_i G(U^{-1}Y(t))$ is the gradient force from free energy. $m_i=k_BT / \langle \dot{Y}^2_i\rangle$ is the effective mass, $K_i$ the memory kernel, and $R_i$ the noise. In practice, when $\langle \dot{X}(t)\dot{X}^T(t)\rangle$ has only negligible off-diagonal terms, we simply let $Y=X$. 


The memory kernel is expanded as  $K_i(s) = \sum_{jl} \phi^{\theta}_{ijl} \theta_{jl}(s) + \phi^{\gamma}_{ijl} \gamma_{jl}(s)$ on a finite set of decay-Fourier basis, 
given by $\theta_{jl}(s)=e^{-\frac{s}{\tau_j}}\cos(\frac{2\pi (l-1) s}{L\tau_j})$ and $\gamma_{jl}(s)=e^{-\frac{s}{\tau_j}}\sin(\frac{2\pi (l-1) s}{ L\tau_j})$, with $1 \leq j\leq J$ and $1 \leq l\leq L$.  $\phi^{\theta/\gamma}$  are parameters. $J$ and $L$ are fixed integers, preferably chosen to be smaller than ten.  $J$ and $L$ control the size of the basis. The decay-Fourier basis characterizes the short-term oscillation and long-term decaying at the same time. In particular, $\tau = (\tau_1,\cdots, \tau_J)$ (arranged in ascending order) will be optimized to fit the multi-scale relaxation time of the memory. The shortest timescale here is typically of the order of one picosecond (ps), associated with local molecular vibration. Meanwhile, the longest timescale is associated with slow conformational changes, hence unbounded. In practice, however, the consideration should be limited to ns scale due to finite and noisy all-atom MD data. Capturing memory effects on a longer timescale should be avoided by adopting a better CG representation of a molecular system. Therefore, we limit the goal of AIGLE to recovering the CG variables's diffusion pattern on the ps$\sim$ns scale.

To be consistent with the memory kernel under the constraint of the 2FDT, the noise $R_i(t)$ is constructed with colored noises obtained from filtering arbitrary Gaussian white noise signal $w_i$ with the same decay-Fourier basis. The colored noises are defined as $\chi^{\theta}_{ijl}(t)=\int_0^\infty \theta_{jl}(s) w_i(t-s) ds$ and $\chi^{\gamma}_{ijl}(t)=\int_0^\infty \gamma_{jl}(s) w_i(t-s) ds$. The total noise acting on $Y_i$ is then $R_i(t)=\sum_{jl} \sigma^{\theta}_{ijl}\chi^{\theta}_{ijl}(t) +  \sigma^{\gamma}_{ijl}\chi^{\gamma}_{ijl}(t)$.  $\sigma^{\theta/\gamma}$ are parameters. 
The 2FDT requires $k_BTK_i(s)=-\langle R_i(t+s)R_i(t) \rangle$ for sufficiently large $t$, leading to exact analytic expressions of $\phi^{\theta/\gamma}$ in terms of $\tau$ and $\sigma^{\theta/\gamma}$(see Methods). Therefore, the free parameters in $K$ and $R$ are reduced to only $\tau$ and $\sigma^{\theta/\gamma}$.  

These parameters are learned from trajectory data of $Y(t)$ through a variational determination of the optimal $K$. In a nutshell, given predetermined $G$ and a trial set of $\tau$ and $\sigma^{\theta/\gamma}$, a trial $K$ is fixed. A time series of ``physical'' noise, $R^{*}_i(t)$, can be extracted by inverting Eq.~(\ref{gle}).  The goal is to enforce for all $x\geq 0$ the orthogonality between the ``physical'' noise and system velocity, given as $\langle\sqrt{m_i}R^{*}_i(x)v_i(0) \rangle=0$. The orthogonality condition is a fundamental requirement for simulating AIGLE with machine-generated noise $R_i$ that mimics the ``physical'' noise while not keeping memory of the physical system. The orthogonality condition can be transformed into $\langle  (\dot{P_i}(x) - F_i(x))v_i(0) \rangle = \int_0^x  m_iK_i(x-s)\langle v_i(s)v_i(0) \rangle  ds$. Integrating both sides of the equation from $x=0$ to $x=t>0$, one obtains 
\begin{equation}\label{orthog}
\small
 \langle  (P_i(t)-P_i(0)- I_i(t))v_i(0) \rangle = m_i\int_0^t  K_i(t-s) D_i(s) ds.
 \end{equation}
Here, $I_i(t)=\int_0^t F_i(s)ds$ is the impulse from the gradient force. 
The dynamical diffusivity, $D_i(s)=\int_0^s \langle v_i(x)v_i(0) \rangle dx$, can be calculated as half the time derivative of the mean squared displacement (MSD) $\Delta_i(s)=\langle |Y_i(s)-Y_i(0)|^2 \rangle$. 

Based on Eq.~(\ref{orthog}), we propose a variational principle for extracting, from MD data, the optimal $\tau$ and $\sigma^{\theta/\gamma}$ that try to preserve both short-term and long-term diffusion pattern of the CG variables. The details are given in Methods.
After the determination of optimal parameters, the simulation of the GLEs can be parallelized as a special case of extended Markovian dynamics~\cite{ceriotti2010colored}. Details of the integration scheme are also in Methods. Last, the AILE is the Markovian limit of AIGLE, given by $\dot{P_i}(t) = F_i(t) - \eta_i P_i(t) + \sqrt{m_i}r_i(t)$.
$\eta_i = -\int_0^\infty K_i(s) ds$. The closed-form expression for $\eta_i$ is given in the supplementary information (SI). The white noise, $r_i$, is directly determined by the Markovian 2FDT.

\subsection{When is AIGLE appropriate for enforcing dynamical consistency}\label{sec:when}
With Eq.~(\ref{orthog}), the question of ``when'' GLE can enforce DC is projected to the existence of a $K_i$ satisfying Eq.~(\ref{orthog}), provided $\tau_J$ is finite. 

We first consider the case of bounded CG variables that $\lim\limits_{s\rightarrow\infty}D_i(s)=0$. A finite $\tau_J$ then requires $\lim\limits_{s\rightarrow\infty}\langle (P_i(t)-P_i(0)- I_i(t))v_i(0) \rangle=0$. Considering velocity autocorrelation eventually vanishes, we obtain $\lim\limits_{s\rightarrow\infty} \langle I_i(t)v_i(0) \rangle =-k_BT$. Let $\zeta_i(t) = -\beta \langle I_i(t)v_i(0) \rangle - 1$. $\zeta_i(t)$ can be estimated from all-atom MD data without the notion of memory or noise. 
We therefore use the large-$t$ behavior of $\zeta_i(t)$ as a practical measure for how well GLE is expected to enforce DC for a CG model. $\|\zeta_i(t)\| \ll 1$ for $t\gg 1$ps and all $i$ is a rule of thumb for appropriate GLE description of CG dynamics.
The physical meaning of obscurely defined $\zeta_i$ becomes obvious when we consider the simple harmonic system $F_i(t)=- c Y_i(t)$, where $\zeta_i(t)$ can be reduced to $-\beta c \langle Y_i(t)Y_i(0)\rangle$. Here, a non-vanishing $\zeta_i(t)$ at the large-$t$ limit means long-lasting position-position correlation, which may connect to the dynamical caging effects~\cite{tuckerman2023statistical}. Dynamic caging refers to a sluggish bath that keeps an infinitely long memory of a physical system. In practice, this may suggest an inappropriate choice of CG variables for describing CG dynamics, instead of a need to extend $\tau_J$ to an unphysically large timescale. 

Then, we consider the case of unbounded CG variables with $D_i(t)$ converging to the normal diffusion constant $D_i^\infty$ when $t\rightarrow \infty$. 
Finite $\tau_J$ requires  $\lim\limits_{s\rightarrow\infty}\langle (P_i(t)-P_i(0)- I_i(t))v_i(0) \rangle= -m_i D_i^\infty \eta_i$, leading to $\lim\limits_{s\rightarrow\infty}\zeta_i = -\beta m_i D_i^\infty \eta_i < 0$. Here, the large-$t$ limit of $\zeta_i$ is connected to the friction constant of AILE, which is not intrinsic to all-atom MD. Calculating $\eta_i$ from $\zeta_i(t)$ is also unreliable for slow diffusion due to accumulated error in $I_i(t)$.  However, the large-$t$ limit of $\zeta_i$ can serve as a rough consistency check after the parameterization of AIGLE. 

These requirements on $\zeta_i(t)$ serve as weak conditions for the applicability of GLE. A stronger requirement is raised by Refs.~\cite{darve2001calculating, noid2008multiscale} for using fixed effective mass and memory kernel in a CG model: the generalized mass matrix $M(q)$, given as $M^{-1}(q)=\sum_k \frac{1}{m^{\mathrm{atom}}_k}\frac{\partial f_X(q)}{\partial q_k}\frac{\partial f_X^T(q)}{\partial q_k}$ ($m^{\mathrm{atom}}_k$ is the mass of the atom $q_k$ is associated with), should be independent of the atomic position $q$. Non-linear CVs typically do not satisfy this requirement. However, fixed effective mass and memory kernel can still be a decent approximation for some non-linear CVs if the eliminated degrees of freedom have a similar dynamical correlation in different metastable states.
To overcome this limitation, GLEs with position-dependent mass or memory kernel have been explored~\cite{hummer2005position, lee2019}. They are out of the scope of this paper for practicality in high-dimensional cases. 

Case studies of a toy polymer and an alanine dipeptide molecule are reported below. The chosen CG variables meet the proposed requirement on $\zeta$. The requirement is however not satisfied when we try to coarse-grain the FiP-35 ww domain protein, one of the most intensively studied small proteins, with RMSD of its two hairpin components as two-dimensional CVs~\cite{a2012dominant}: $|\zeta(t)|$ does not drop near zero on the dataset, suggesting long-lasting memory effects. This points to the difficulty in reaching DC with phenomenological CG variables and the outlook of systematic optimization of CG variables for long-term DC. Some CV identification algorithms, such as time-lagged independent component analysis ~\cite{schwantes2013improvements, perez2013identification} or the variational approach for Markov processes ~\cite{wu2020variational}, can isolate slow collective motions while filtering out fast dynamics, making them potential candidates for constructing CVs with reduced memory effects. 

\subsection{Particle-based coarse-graining}\label{sec:polymer}

\begin{figure}[tb]
    \centering
    \includegraphics[width=\linewidth]{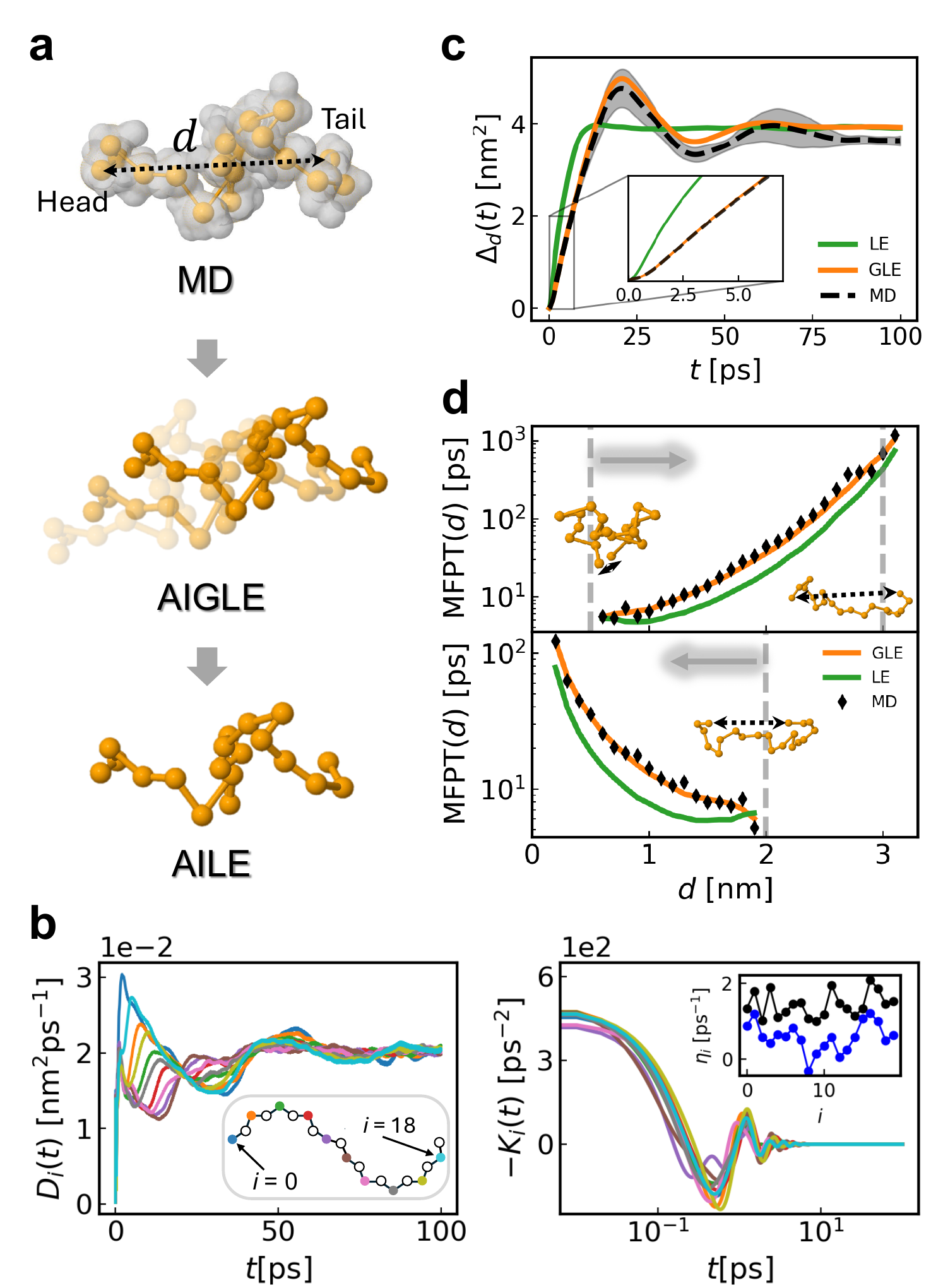}
    \caption{(a) Sketches of the all-atom model and the CG models of the toy polymer. The grey shades in the all-atom model represent the dangling particles attached to the backbone atoms. (b) Left: The dynamical diffusivity as a function of time. Right: The memory kernel as a function of time. The inset plots the effective Markovian friction $\eta_i$ for all backbone atoms. (c) MSD of $d$ as a function of time.  (d) MFPT as a function of $d$ for expanding (upper) and contracting (lower) processes.}
    \label{fig:polymer}
\end{figure}

We study a toy polymer model (Fig.~\ref{fig:polymer}(a)), where the backbone is a chain of 20 identical particles connected by harmonic springs of the elastic constant $k=1000 \mathrm{kJ/(mol\cdot nm^2)}$ and the equilibrium length $l_0=0.3$nm. Each backbone particle has the mass of a Carbon atom (12 Dalton) and is bonded exclusively with five dangling particles through the same harmonic springs described by $k$ and $l_0$. The whole polymer contains 120 particles. For each backbone atom respectively, the associated dangling particles are assigned with random masses sampled uniformly from $[1,20]$ Dalton. So the polymer chain becomes heterogeneous. 
We obtain all-atom MD trajectories of the toy polymer at temperature $T=300$K by coupling the Langevin thermostat to each particle with a relaxation time of 10ps. The dynamical diffusivity, $D_i(t)$, are plotted in Fig.~\ref{fig:polymer}(b, left) for half of the backbone particles ($i=0,2,\cdots, 18$). It appears that backbone particles closer to the ends of the polymer diffuse faster within $t<50$ps, the timescale of anomalous diffusion. All $D_i(t)$ converges to roughly $0.02 \mathrm{nm^2ps^{-1}}$ for $T>50$ps, representing the global normal diffusion rate $D^\infty$ of the polymer. $\zeta_i(t)$  saturates around $t=100$ps, leading to a rough estimation of $\eta_i$ plotted in the inset of Fig.~\ref{fig:polymer}(b, right) as blue dots.

We use coordinates of backbone atoms as CG variables (Fig.~\ref{fig:polymer}(a)) of the polymer. 
Orthogonal transformation of CG variables is not performed because the cross-velocity correlation between backbone atoms is negligible. 
Because the system is harmonic, the free energy model is exact and the GLE description of CG dynamics can be accurate. The effective interaction of the CG particles (mass unchanged) is the same harmonic springs on the backbone. 
Then, we derive AIGLE and AILE for CG dynamics (see SI for details). For AIGLE, we choose $I=3$, $J=4$. Trained on $10$ns-long all-atom MD trajectory, the AIGLE converges to $\tau=(0.37, 0.92, 1.26)$ps by imposing Eq.~(\ref{orthog}) for $t<6$ps (see SI for details). $\tau$ lies in the range of vibrational frequencies of dangling particles.  The memory kernels $K_i(t)$ are plotted in Fig.~\ref{fig:polymer}(b, right) for $i=0,2,\cdots, 18$. The corresponding $\eta_i$ obtained from integrating $K_i(t)$ is plotted in the inset as black dots. The $\eta_i$ from AIGLE is of the same order as $\eta_i$ estimated from $\zeta_i$. Note that unphysical $\eta_i<0$ appears from $\zeta_i$ due to error accumulated in $\zeta_i(t)$. In contrast, $\eta_i$ from AIGLE is always physical. 

AIGLE and AILE simulate CG dynamics for $100$ns after $100$ps equilibration at $T=300$K. Comparing them to all-atom MD, we study the dynamical behaviors of $d$, the end-to-end distance of the polymer (see Fig.~\ref{fig:polymer}(a)).
The MSD of $d$, denoted by $\Delta_d(t)$, is plotted in Fig.~\ref{fig:polymer}(c). The $\Delta_d(t)$ from AIGLE and AILE is converged. The $\Delta_d(t)$ computed from 10ns-long MD trajectory contains non-negligible uncertainty, indicated as grey shadows. AIGLE agrees closely with MD on anomalous diffusion behavior within the entire region of $t<75$ps, while AILE suppresses the anomalous diffusion into the region $t<10$ps. The discrepancy between AILE and AIGLE is also significant in the short term. The inset of Fig.~\ref{fig:polymer}(c) shows a perfect agreement of AIGLE and MD on the order of 1ps, while AILE deviates from MD from the beginning. 

A major consequence of deviated $\Delta_d(t)$ is the deviation in mean first-passage time (MFPT) between characteristic system states.  Fig.~\ref{fig:polymer}(c) shows respectively the MFPT associated with the expansion of the polymer (upper panel) from $d=0.5$nm to $d=3$nm, and the contraction of the polymer (lower panel) from $d=2.5$nm to $d=0.5$nm. In both cases, AIGLE is accurate, while AILE underestimates MFPT several times. Moreover, a more significant error in MFPT will be present if one uses an arbitrary $\eta_i$ for LE, which is already analyzed by Refs.~\cite{ayaz2021non, dalton2023fast} for 1D CG models. 

\begin{figure*}[tb]
    \centering
    \includegraphics[width=\linewidth]{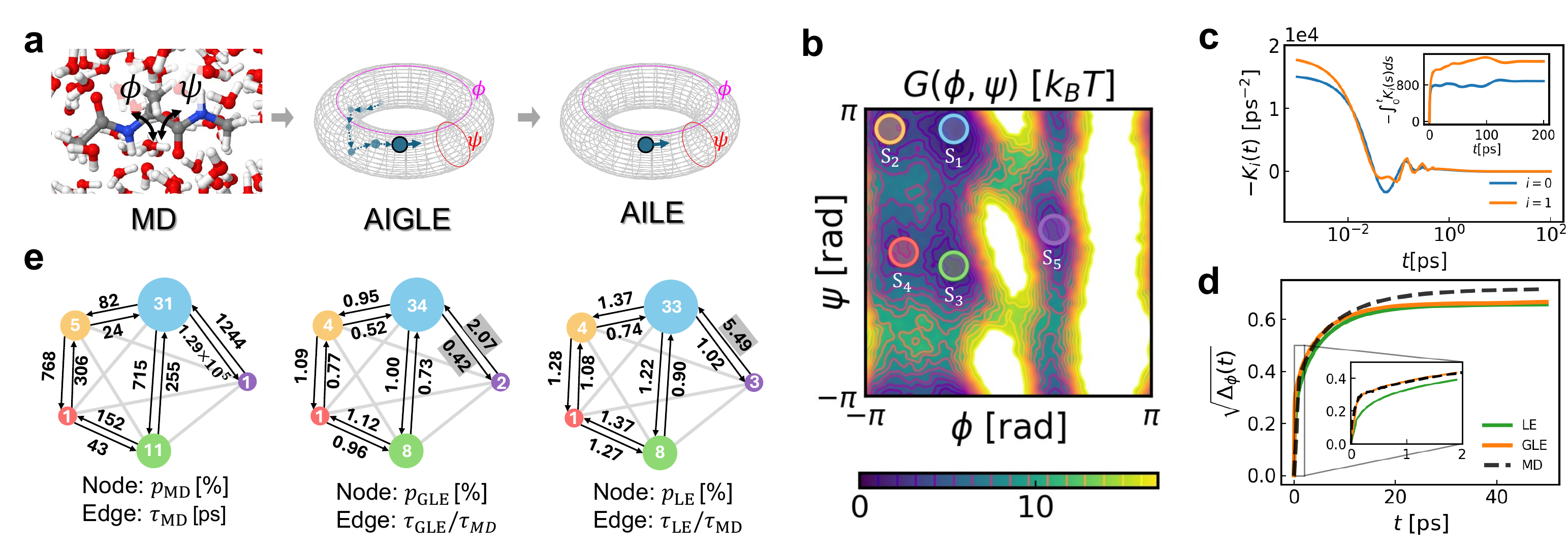}
    \caption{(a) Schematic representations of all-atom MD, AIGLE, and AILE simulation of alanine dipeptide. (b) The heatmap of the free energy surface $G(\phi, \psi)$. Colored solid circles mark five representative states. (c) Memory kernels of the orthogonally transformed CVs. The inset shows the integrated memory kernels. (d) RMSD of $\phi$ as a function of time. (e)  Graph representation of $\{\mathrm{S_1, \cdots, S_5}\}$, matching panel (b) by color. The size of the nodes is arranged in ascending order of the equilibrium probability, reported as node labels. The MFPTs are reported as labels of bidirectional edges. We highlight the edge label with grey shade when $\frac{\tau_{\mathrm{GLE}}}{\tau_{\mathrm{MD}}}$ or $\frac{\tau_{\mathrm{LE}}}{\tau_{\mathrm{MD}}}$ is larger than 2 or smaller than 0.5. }
    \label{fig:alanine}
\end{figure*}

Considering AIGLE and AILE give the same, static $\Delta_d(t)$ in the long term, the discrepancy in MFPT originates in anomalous diffusion of individual backbone atoms, specifically their large initial diffusion rate revealed by Fig.~\ref{fig:polymer}(b), which causes the overshooting of $\Delta_d(t)$ at $t=25$ps. GLE can capture the overshooting by mimicking impulses from dangling particles through memory and noise, while LE can not. 

This case study elucidates ``why'' GLE is preferred over LE by showing excellent accuracy of GLE on dynamical properties that are relevant to practical molecular modeling of realistic polymers.

\subsection{Collective variable-based coarse-graining}

Here, the all-atom MD model is an alanine dipeptide molecule dissolved in 863 water molecules at $T=300$K, described by the AMBER14~\cite{maier2015ff14sb} force field and the TIP3P-FB~\cite{wang2014building} water model. The dihedral angles, $\phi$ and $\psi$ (see Fig.~\ref{fig:alanine}(a)), are chosen as CVs. The free energy surface $G(\phi, \psi)$ of the CVs is obtained from 20ns well-tempered metadynamics simulation and plotted in Fig.~\ref{fig:alanine}(b), where five representative metastable basins, $\{\mathrm{S_1, \cdots, S_5}\}$, are marked with colored circles of radius 0.3rad. Then, a 60ns unbiased, continuous (not mapped back to the minimal image), all-atom NVT-MD trajectory is generated for training AIGLE with I=5 and J=4. Note that this unbiased trajectory does not visit the $S_5$ basin. Non-ergodic unbiased exploration is typical, if not always, in practice. We will evaluate critically the performance of AIGLE/AILE under this circumstance. 

We perform orthogonal transformation $(Y_1, Y_2)^T=U (\phi,\psi)^T$ due to considerable cross velocity correlation between $\phi$ and $\psi$. $\|\zeta_i(t)\|$ (associated to $Y_i$) is found to oscillate within 0.2 for $t>50$ps. $\|\zeta_i(t)\|$ does not converge to zero because $\psi$ is drifting. The oscillation of $\zeta_i(t)$ suggests a minor long-lasting memory effect.
Trained on the trajectory of $Y$, AIGLE converges to $\tau=(0.06, 0.12, 2.37, 11.04, 33.04)$ps by imposing Eq.~(\ref{orthog}) for $t<100$ps (see SI for details). The memory kernels $K_i(t)$ are shown in Fig.~\ref{fig:alanine}(c) for $t<1$ps. The long tail of $K_i(t)$ is reflected by $\int_0^t K_i(s)ds$ plotted in the inset. The resultant $\eta_i\approx 1000\mathrm{ps^{-1}}$ causes overdamping.

We use AIGLE, AILE, and all-atom MD to generate respectively 20$\mathrm{\mu s}$, 20$\mathrm{\mu s}$, and 10$\mathrm{\mu s}$ ergodic, continuous trajectories of the CVs. RMSD of $\phi$, denoted by $\sqrt{\Delta_\phi(t)}$, is plotted in Fig.~\ref{fig:alanine}(d). Then, in Fig.~\ref{fig:alanine}(e), we report the equilibrium probability ($p_{\mathrm{MD/GLE/LE}}$) and the MFPT ($\tau_{\mathrm{MD/GLE/LE}}$) associated with $\{\mathrm{S_1, \cdots, S_5}\}$, for all-atom MD/AIGLE/AILE respectively.
AIGLE agrees almost perfectly with all-atom MD on RMSD for the fast-diffusion regime $t\leq 10$ps, while AILE deviates slightly from MD from the beginning (see the inset of Fig.~\ref{fig:alanine}(d, upper)). 
The fast diffusion is attributed to the oscillation of CVs around a metastable valley, caused by fast molecular vibrational modes, and also frequent transitions (see Fig.~\ref{fig:alanine}(e)) between $\mathrm{S_1}$ ($\mathrm{S_3}$) and $\mathrm{S_2}$ ($\mathrm{S_4}$). The regime of $t>10$ps yields instead a slow growth of RMSD on the timescale of nanoseconds, mainly associated with transitions among metastable states. 
AIGLE and AILE slightly underestimate the RMSD for this regime, likely due to a minor error in $G(\phi,\psi)$. Non-negligible error in $G(\phi,\psi)$ is evidenced by the fact that the equilibrium probability of $S_1$ is overestimated by $10\%$ by CG models.

Next, we compare all-atom MD, AIGLE, and AILE on MFPTs. Based on Fig.~\ref{fig:alanine}(e), the overall agreement among the three methods is ensured for transitions within the $\phi < 0$ regime, where both $\frac{\tau_{\mathrm{GLE}}}{\tau_{\mathrm{MD}}}$ and $\frac{\tau_{\mathrm{LE}}}{\tau_{\mathrm{MD}}}$ are controlled within $[0.5,2]$. Although AILE always predicts larger MFPT than AIGLE due to its overdamping nature, the resultant difference is insignificant for the $\phi < 0$ regime. However, for the transition between $\mathrm{S_1}$ to $\mathrm{S_5}$, both AIGLE and AILE deviate from MD considerably. In particular, AIGLE makes the forward hopping from $S_1$ to $S_5$ easier and the backward hopping more difficult. Errors in $G(\phi,\psi)$ may contribute to the discrepancy by lowering the transition barrier and the free energy of $\mathrm{S_5}$.
The latter is evidenced by the overestimated $p_{\mathrm{GLE}}\approx 2.2\%$ associated with $S_5$, in constrast to $p_{\mathrm{MD}}\approx 0.7\%$. With the same free energy surface, AILE is accurate on forward hopping, however, it overestimates the backward MFPT by roughly five times. The hindered backward hopping leads to insufficient equilibration and a $p_{\mathrm{LE}}\approx 3.4\%$ associated to $\mathrm{S_5}$. Likely, the friction in AILE is overestimated for describing the oscillation of CVs in $\mathrm{S_5}$ basin. AIGLE alleviates this issue by having a continuous memory kernel that effectively cancels out friction through convoluting over oscillatory trajectories. Such cancellation however does not address the fundamental difficulty of CG modeling with empirical CVs: the memory and noise differ for different metastable basins. This problem can not be solved by further training GLE on long, ergodic MD trajectories. Optimization of CVs should be involved.

Here, the advantage of AIGLE over AILE is not significant. The CG dynamics appear overdamped for $\phi <0$, reducing the importance of non-Markovian memory and noise effects. Overall, both AIGLE and AILE give access to a relatively accurate representation of the all-atom system with 
more than two orders of magnitude speedup over direct MD simulation.


\section{Conclusion}
We have introduced data-driven approaches AIGLE/AILE for practical learning of GLE and LE models from multi-dimensional, heterogeneous trajectory data of CG variables. AIGLE balances accuracy and efficiency. AILE targets at less accurate but more efficient simulation. 
We have proposed practical criteria for estimating the ability of AIGLE to enforce DC in the long term, without training the models. The criteria are also relevant to optimizing CG variables for DC. It potentially leads to a variational principle that applies to generic anharmonic systems. We leave this to future explorations. 

We have benchmarked the AIGLE/AILE approaches for particle-based and CV-based coarse-graining respectively. For both cases, AIGLE/AILE trained on short trajectories systematically predicts MFPT between arbitrary states without phenomenological theories such as a transition state theory. In the future, AIGLE/AILE may lead to systematic parameterization of more coarse-grained Markov state models where atomistic details are removed, in pursuit of an efficient description of hopping among characteristic/meta-stable states. 

Specifically, for the toy polymer model, AIGLE faithfully predicts the MFPT for long-term, global conformational changes, while AILE is less accurate. For this simple system, the computational cost of integrating AIGLE is comparable to the evaluation of gradient forces. The overhead from modeling non-Markovian effects should become marginal when the force field is more realistic by including long-range non-bonded interaction. AIGLETools interfaces with the MD software OPENMM~\cite{eastman2023openmm} to facilitate future investigation in particle-based CG. 

For alanine dipeptide, AIGLE and AILE demand significantly less computational cost than all-atom MD. Trained on nanoseconds-level data of dihedral angles, they allow relatively accurate descriptions of rare events on the microsecond scale. As demonstrated by Ref.~\cite{shaffer2016enhanced}, advanced enhanced sampling techniques can solve the high-dimensional free energy surface for tens of dihedral angles associated with a protein backbone. Combined with these techniques, high-dimensional AIGLE may consistently simulate microseconds-level backbone dynamics based on nanoseconds-level all-atom MD data. 

Further application of CV-based AIGLE/AILE in generic conformational dynamics problems also requires efforts to optimize the low-dimensional CV, taking full consideration of the requirements from DC in the general, anharmonic setting.

\section*{Methods}
\subsection{Second fluctuation-dissipation theorem on decay-Fourier basis}
Here we derive the exact expression of $\phi^{\theta/\gamma}_{ijl}$, in terms of $\tau$ and $\sigma^{\theta/\gamma}$, as prescribed by the second fluctuation-dissipation theorem.
Expanding the autocorrelation function of the noise with the decay-Fourier basis, we arrive at 
\begin{equation}
\small
\begin{split}
    \langle R_i(t+s)R_i(t) \rangle &= \Big\langle \sum_{jlj'l'}( \sigma^{\theta}_{ijl}\chi^{\theta}_{ijl}(t+s) +  \sigma^{\gamma}_{ijl}\chi^{\gamma}_{ijl}(t+s)) \\
&\times (\sigma^{\theta}_{ij'l'}\chi^{\theta}_{ij'l'}(t) +  \sigma^{\gamma}_{ij'l'}\chi^{\gamma}_{ij'l'}(t)) \Big\rangle
\end{split}
\end{equation}

A straightforward but lengthy derivation leads to 
\begin{equation}\label{eq:chi_corr}
\small
\begin{cases}
    \langle\chi^{\theta}_{ijl}(t+s)\chi^{\theta}_{ij'l'}(t)\rangle =\theta_{jl}(s) M^{\theta\theta}_{jlj'l'}(\tau) - \gamma_{jl}(s) M^{\gamma\theta}_{jlj'l'}(\tau)  \\
    \langle\chi^{\theta}_{ijl}(t+s)\chi^{\gamma}_{ij'l'}(t)\rangle =\theta_{jl}(s) M^{\theta\gamma}_{jlj'l'}(\tau) - \gamma_{jl}(s) M^{\gamma\gamma}_{jlj'l'}(\tau)  \\
    \langle\chi^{\gamma}_{ijl}(t+s)\chi^{\theta}_{ij'l'}(t)\rangle =\theta_{jl}(s) M^{\gamma\theta}_{jlj'l'}(\tau) + \gamma_{jl}(s) M^{\theta\theta}_{jlj'l'}(\tau)  \\
    \langle\chi^{\gamma}_{ijl}(t+s)\chi^{\gamma}_{ij'l'}(t)\rangle =\theta_{jl}(s) M^{\gamma\gamma}_{jlj'l'}(\tau) + \gamma_{jl}(s) M^{\theta\gamma}_{jlj'l'}(\tau).  
\end{cases}
\end{equation}

The $2\times 2 \times J\times L\times J\times L$ tensor $M$ is analytically determined by only $\tau$. The derivation of Eq.~(\ref{eq:chi_corr}) and the exact formula for $M$ is given in SI. 

Matching $K_i(s) = \sum_{jl} \phi^{\theta}_{ijl} \theta_{jl}(s) + \phi^{\gamma}_{ijl} \gamma_{jl}(s)$ and the 2FDT $k_BTK_i(s)=-\langle R_i(t+s)R_i(t) \rangle$, one obtains the final expression for $\phi^{\theta/\gamma}$, given as 
\begin{equation}\label{phi_0}
\small
\begin{split}
    k_BT \phi^{\theta}_{ijl} &= -\sigma_{ijl}^\theta \sum_{j'l'} (\sigma_{ij'l'}^{\theta}M^{\theta\theta}_{jlj'l'}(\tau) + \sigma_{ij'l'}^{\gamma}M^{\theta\gamma}_{jlj'l'}(\tau)) \\
    &- \sigma_{ijl}^\gamma \sum_{j'l'} (\sigma_{ij'l'}^{\theta}M^{\gamma\theta}_{jlj'l'}(\tau) + \sigma_{ij'l'}^{\gamma}M^{\gamma\gamma}_{jlj'l'}(\tau))
\end{split}
\end{equation}
and 
\begin{equation}\label{phi_1}
\small
\begin{split}
    k_BT \phi^{\gamma}_{ijl} &= \sigma_{ijl}^\theta \sum_{j'l'} (\sigma_{ij'l'}^{\theta}M^{\gamma\theta}_{jlj'l'}(\tau) + \sigma_{ij'l'}^{\gamma}M^{\gamma\gamma}_{jlj'l'}(\tau)) \\ 
    &- \sigma_{ijl}^\gamma \sum_{j'l'} (\sigma_{ij'l'}^{\theta}M^{\theta\theta}_{jlj'l'}(\tau) + \sigma_{ij'l'}^{\gamma}M^{\theta\gamma}_{jlj'l'}(\tau)).
\end{split}
\end{equation}

\subsection{Variational principle for orthogonality condition}
 Based on Eq.~(\ref{orthog}), we define the orthogonality loss $\varepsilon_i(t)=  \langle  (P_i(t)- P_i(0)- I(t))v_i(0) \rangle - m_i\int_0^t  K_i(t-s) D_i(s) ds$. 
 Notice $\langle P_i(0)v_i(0)\rangle = k_BT$ and expand $K_i$ over the decay-Fourier basis. We obtain
 \begin{equation}\label{epsilon}
 \begin{split}
     \varepsilon_i(t)=&-k_BT+\langle  (P_i(t)- I(t))v_i(0) \rangle  \\
     &-m_i   \sum_{jl}\left(\phi^{\theta}_{ijl} \tilde{D}_{ijl}^{\theta}(t)  + \phi^{\gamma}_{ijl} \tilde{D}_{ijl}^{\gamma}(t)  \right).
 \end{split}    
 \end{equation}
Here, $\tilde{D}_{ijl}^{\theta}(t)$ represents the decay-Fourier transform $\int_0^t\theta_{jl}(t-s)D_i(s)ds$. Similarly, $\tilde{D}_{ijl}^{\gamma}(t)=\int_0^t\gamma_{jl}(t-s)D_i(s)ds$. $D_i(s)$ is fixed by the data through the finite difference of the MSD of $Y_i$. The term $\langle  (P_i(t)- I(t))v_i(0) \rangle$ is also fixed by averaging over MD data.

Considering the orthogonality loss function is eventually a functional of model parameters $\tau$ and $\sigma^{\theta/\gamma}$ (see Eq.~(\ref{phi_0}-\ref{phi_1})), we define the global objective function
 \begin{equation}\label{loss}
     \mathcal{L}[\tau, \sigma^{\theta/\gamma}] = \frac{1}{T_{\mathrm{cut}}} \sum_i\int_0^{T_{\mathrm{cut}}} \left| \varepsilon_i(t) \right|^2 dt.
 \end{equation}
for enforcing the orthogonality condition for a finite period. The cutoff $T_{\mathrm{cut}}$ should be chosen to the order of the timescale of anomalous diffusion of the CVs. The optimal model parameters are defined as the minimizer of $\mathcal{L}[\tau, \sigma^{\theta/\gamma}]$ based on given MD data. The minimization can be done with a gradient descent algorithm with even-spacing ( or adaptive) discretization of the integration in Eq.~(\ref{loss}) (see SI for details). The convergence is typically rapid, benefitting from the simple quadratic dependence of $\phi^{\theta/\gamma}$
on $\sigma^{\theta/\gamma}$.

\subsection{GLE Simulation with Markovian integrator}
Here we show how to integrate GLE in a Markovian way by introducing auxiliary variables. 
We define the memory force  $F^{\theta}_{ijl}=\int_0^t \theta_{jl}(s)P_i(t-s)ds$ and $F^{\gamma}_{ijl}=\int_0^t \gamma_{jl}(s)P_i(t-s)ds$. 
We then extend the physical phase space $(\{P_i\},\{Y_i\})$ to the auxiliary phase space $(\{P_i\},\{Y_i\}, \{R_i\},\{F^{\theta}_{ijl}\}, \{F^{\gamma}_{ijl}\}, \{\chi^{\theta}_{ijl}\}, \{\chi^{\gamma}_{ijl}\})$. In the auxiliary phase space, the GLEs can be disguised as a set of Markovian dynamical equations:
\begin{equation}
\small
    \begin{cases}
    \dot{P_i}(t) &= -\partial_i G(U^{-1}Y(t)) + \sum_{jl} (\phi^{\theta}_{ijl}F^{\theta}_{ijl}(t) + \phi^{\gamma}_{ijl} F^{\gamma}_{ijl})   \\
    &\ \ \ + \sqrt{m_i}\sum_{jl} (\sigma^{\theta}_{ijl}\chi^{\theta}_{ijl}(t) +  \sigma^{\gamma}_{ijl}\chi^{\gamma}_{ijl}(t)) \\
    \dot{Y}_i(t) &= m_i^{-1} P_i(t)\\

    \dot{F}^{\theta}_{ijl}(t) &= P(t) - \tau_i^{-1} F^{\theta}_{ijl}(t) - \alpha_{jl} F^{\gamma}_{ijl}(t) \\

    \dot{F}^{\gamma}_{ijl}(t) &= - \tau_i^{-1} F^{\gamma}_{ijl}(t) + \alpha_{jl} F^{\theta}_{ijl}(t) \\

    \dot{\chi}^{\theta}_{ijl}(t) &= \omega_i(t) - \tau_i^{-1} \chi^{\theta}_{ijl}(t) - \alpha_{jl} \chi^{\gamma}_{ijl}(t) \\

    \dot{\chi}^{\gamma}_{ijl}(t) &= - \tau_i^{-1} \chi^{\gamma}_{ijl}(t) + \alpha_{jl} \chi^{\theta}_{ijl}(t) \\
    \end{cases}.
\end{equation}
$\alpha_{jl}$ is the abbreviation of $\frac{2\pi (l-1) }{ L\tau_j}$. This set of equations can be integrated with Markovian integrators after discretization. Note that the time derivative of an auxiliary variable depends only on auxiliary variables with the same subscripts. So the complexity of integrating auxiliary variables is O($nJL$). Since $J$ and $L$ are typically integers smaller than 10, the additional cost introduced by including non-Markovian effects is linear scaling to the dimensionality of the CVs. 

\section*{Data availability}
The package AIGLETools and the Jupyter Notebooks for reproducing results in this work are publicly available at ~\cite{aigletools}.

\section*{Acknowledgments}
We thank Roberto Car for fruitful discussions.
P.X. was supported by the Computational Chemical Sciences Center:  Chemistry in Solution and Interfaces (CSI) funded by DOE Award DE-SC0019394. P.X. and W.E were also supported by a gift from iFlytek to Princeton University. W.E was supported by the Basic Science Center of National Natural Science Foundation of China with Award NSFC No.12288101. The work reported in this paper was substantially performed using the Princeton Research Computing resources at Princeton University which is consortium of groups led by the Princeton Institute for Computational Science and Engineering (PICSciE) and Office of Information Technology's Research Computing.
\bibliography{aigle_bio}
\bibliographystyle{naturemag}

\clearpage
\onecolumngrid





\setcounter{section}{0}
\setcounter{subsection}{0}
\setcounter{equation}{0}

\section*{Supplementary Information}

\subsection{Derivation of second fluctuation-dissipation theorem on decay-Fourier basis}

In the main text, we have derived the autocorrelation
function of the noise with the decay-Fourier basis, given as
\begin{equation}\label{si: eq1}
\begin{split}
    \langle R_i(t+s)R_i(t) \rangle &= \sum_{jlj'l'}\Big( 
       \sigma^{\theta}_{ijl}\sigma^{\theta}_{ij'l'}\langle\chi^{\theta}_{ijl}(t+s)\chi^{\theta}_{ij'l'}(t)\rangle + \sigma^{\theta}_{ijl}\sigma^{\gamma}_{ij'l'}\langle\chi^{\theta}_{ijl}(t+s)\chi^{\gamma}_{ij'l'}(t)\rangle \\       
    &+  \sigma^{\gamma}_{ijl}\sigma^{\theta}_{ij'l'}\langle\chi^{\gamma}_{ijl}(t+s)\chi^{\theta}_{ij'l'}(t)\rangle +\sigma^{\gamma}_{ijl}\sigma^{\gamma}_{ij'l'}\langle\chi^{\gamma}_{ijl}(t+s)\chi^{\gamma}_{ij'l'}(t)\rangle \Big).
\end{split}
\end{equation}
Recall that $\chi^{\theta}_{ijl}(t)=\int_0^\infty \theta_{jl}(s) w_i(t-s) ds$ and $\chi^{\gamma}_{ijl}(t)=\int_0^\infty \gamma_{jl}(s) w_i(t-s) ds$.
We have
\begin{equation}\label{si:eq2}
\begin{split}
    \langle\chi^{\theta}_{ijl}(t+s)\chi^{\theta}_{ij'l'}(t)\rangle &= \Big\langle
    \int_0^\infty e^{-\frac{s'}{\tau_j}}\cos(\tfrac{2\pi (l-1) s'}{L\tau_j}) \omega_i(t+s-s')ds' 
    \int_0^\infty  e^{-\frac{s''}{\tau_{j'}}}\cos(\tfrac{2\pi (l'-1) s''}{L\tau_{j'}}) \omega_i(t-s'')ds'' \Big\rangle \\
     &=\int_0^\infty\int_0^\infty e^{-\frac{s'}{\tau_j}}\cos(\tfrac{2\pi (l-1) s'}{L\tau_j}) 
      e^{-\frac{s''}{\tau_{j'}}}\cos(\tfrac{2\pi (l'-1) s''}{L\tau_{j'}})   \delta(s-s'+s'') ds'ds'' \\
     &=\int_0^\infty  e^{-\frac{s+s''}{\tau_j}}\cos(\tfrac{2\pi (l-1) (s+s'')}{L\tau_j}) 
      e^{-\frac{s''}{\tau_{j'}}}\cos(\tfrac{2\pi (l'-1) s''}{L\tau_{j'}})    ds'' \\
      &=e^{-\frac{s}{\tau_j}} \int_0^\infty  e^{-\frac{s''}{\tau_j} -\frac{s''}{\tau_{j'}}}\cos(\tfrac{2\pi (l-1) (s+s'')}{L\tau_j}) 
       \cos(\tfrac{2\pi (l'-1) s''}{L\tau_{j'}})    ds'' \\
    &=e^{-\frac{s}{\tau_j}} \cos(\tfrac{2\pi (l-1) s}{L\tau_j})  \int_0^\infty  e^{-\frac{s''}{\tau_j} -\frac{s''}{\tau_{j'}}} \cos(\tfrac{2\pi (l-1)s''}{L\tau_j}) 
       \cos(\tfrac{2\pi (l'-1) s''}{L\tau_{j'}})    ds'' \\
    & -e^{-\frac{s}{\tau_j}} \sin(\tfrac{2\pi (l-1) s}{L\tau_j})  \int_0^\infty  e^{-\frac{s''}{\tau_j} -\frac{s''}{\tau_{j'}}} \sin(\tfrac{2\pi (l-1)s''}{L\tau_j}) 
       \cos(\tfrac{2\pi (l'-1) s''}{L\tau_{j'}})    ds''
\end{split}
\end{equation}

Let 
\begin{equation}\label{si:m}
    \begin{cases}
        M^{\theta\theta}_{jlj'l'}(\tau)=\int_0^\infty  e^{-\frac{s}{\tau_j} -\frac{s}{\tau_{j'}}} \cos(\tfrac{2\pi (l-1)s}{L\tau_j}) 
       \cos(\tfrac{2\pi (l'-1) s}{L\tau_{j'}})    ds \\
       M^{\theta\gamma}_{jlj'l'}(\tau)=\int_0^\infty  e^{-\frac{s}{\tau_j} -\frac{s}{\tau_{j'}}} \cos(\tfrac{2\pi (l-1)s}{L\tau_j}) 
       \sin(\tfrac{2\pi (l'-1) s}{L\tau_{j'}})    ds \\
       M^{\gamma\theta}_{jlj'l'}(\tau)=\int_0^\infty  e^{-\frac{s}{\tau_j} -\frac{s}{\tau_{j'}}} \sin(\tfrac{2\pi (l-1)s}{L\tau_j}) 
       \cos(\tfrac{2\pi (l'-1) s}{L\tau_{j'}})    ds \\
      M^{\gamma\gamma}_{jlj'l'}(\tau)=\int_0^\infty  e^{-\frac{s}{\tau_j} -\frac{s}{\tau_{j'}}} \sin(\tfrac{2\pi (l-1)s}{L\tau_j}) 
       \sin(\tfrac{2\pi (l'-1) s}{L\tau_{j'}})    ds.
    \end{cases}
\end{equation}
Let $\lambda_j=1/\tau_j$ and $\alpha_{jl}=\tfrac{2\pi (l-1)}{L\tau_j}$.
Eq.~(\ref{si:m}) can be transformed into
\begin{equation}
    \begin{cases}
        M^{\theta\theta}_{jlj'l'}(\tau)=\frac{\lambda_j + \lambda_{j'}}{2} \left( \frac{1}{(\lambda_j + \lambda_{j'})^2+(\alpha_{jl}-\alpha_{j'l'})^2} + \frac{1}{(\lambda_j + \lambda_{j'})^2+(\alpha_{jl}+\alpha_{j'l'})^2}
  \right) \\
       M^{\theta\gamma}_{jlj'l'}(\tau)=\frac{1}{2} \left( \frac{-(\alpha_{jl}-\alpha_{j'l'})}{(\lambda_j + \lambda_{j'})^2+(\alpha_{jl}-\alpha_{j'l'})^2} + \frac{\alpha_{jl}+\alpha_{j'l'}}{(\lambda_j + \lambda_{j'})^2+(\alpha_{jl}+\alpha_{j'l'})^2}
  \right) \\
       M^{\gamma\theta}_{jlj'l'}(\tau)=\frac{1}{2} \left( \frac{(\alpha_{jl}-\alpha_{j'l'})}{(\lambda_j + \lambda_{j'})^2+(\alpha_{jl}-\alpha_{j'l'})^2} + \frac{\alpha_{jl}+\alpha_{j'l'}}{(\lambda_j + \lambda_{j'})^2+(\alpha_{jl}+\alpha_{j'l'})^2}
  \right)  \\
      M^{\gamma\gamma}_{jlj'l'}(\tau)=\frac{\lambda_j + \lambda_{j'}}{2} \left( \frac{1}{(\lambda_j + \lambda_{j'})^2+(\alpha_{jl}-\alpha_{j'l'})^2} - \frac{1}{(\lambda_j + \lambda_{j'})^2+(\alpha_{jl}+\alpha_{j'l'})^2}
  \right).
    \end{cases}
\end{equation}

Eq.~(\ref{si:eq2}) can rewritten as 
\begin{equation}
    \langle\chi^{\theta}_{ijl}(t+s)\chi^{\theta}_{ij'l'}(t)\rangle = \theta_{jl}(s)M^{\theta\theta}_{jlj'l'}(\tau) - \gamma_{jl}(s)M^{\gamma\theta}_{jlj'l'}(\tau).
\end{equation}

Similarly, we can also derive 
\begin{equation}
    \langle\chi^{\theta}_{ijl}(t+s)\chi^{\gamma}_{ij'l'}(t)\rangle =\theta_{jl}(s) M^{\theta\gamma}_{jlj'l'}(\tau) - \gamma_{jl}(s) M^{\gamma\gamma}_{jlj'l'}(\tau),
\end{equation}
\begin{equation}
    \langle\chi^{\gamma}_{ijl}(t+s)\chi^{\theta}_{ij'l'}(t)\rangle =\theta_{jl}(s) M^{\gamma\theta}_{jlj'l'}(\tau) + \gamma_{jl}(s) M^{\theta\theta}_{jlj'l'}(\tau),
\end{equation}
and
\begin{equation}
    \langle\chi^{\gamma}_{ijl}(t+s)\chi^{\gamma}_{ij'l'}(t)\rangle =\theta_{jl}(s) M^{\gamma\gamma}_{jlj'l'}(\tau) + \gamma_{jl}(s) M^{\theta\gamma}_{jlj'l'}(\tau).
\end{equation}

Then, matching $K_i(s) = \sum_{jl} \phi^{\theta}_{ijl} \theta_{jl}(s) + \phi^{\gamma}_{ijl} \gamma_{jl}(s)$ and $k_BTK_i(s)=-\langle R_i(t+s)R_i(t) \rangle$ will give the final expressions for 
$\phi^{\theta/\gamma}_{ijl}$ (Eqs.(5-6) in the main text) that impose the exact second fluctuation-dissipation theorem. 

\subsection{Objective function}
The objective function (Eq.(8) in the main text) is
 \begin{equation}
     \mathcal{L}[\tau, \sigma^{\theta/\gamma}] = \frac{1}{T_{\mathrm{cut}}} \sum_i\int_0^{T_{\mathrm{cut}}} \left| \varepsilon_i(t) \right|^2 dt.
 \end{equation}

Discretizing it with the time step $\delta t$ of an MD trajectory $Y(t)$ leads to the loss function
 \begin{equation}
     \mathcal{L}_Y[\tau, \sigma^{\theta/\gamma}] = \frac{\delta t}{T_{\mathrm{cut}}} \sum_i \sum_{n=1}^N \left| \varepsilon_i(n\delta t) \right|^2 ,
 \end{equation}
where $N$ is the closest integer to $T_{\text{cut}}/\delta t$.
The orthogonality loss becomes
\begin{equation}\label{epsilon}
     \varepsilon_i(n\delta t)=-k_BT+m_i C_i^{vv}(n\delta t)- C_i^{Iv}(n\delta t) 
     -m_i   \sum_{jl}\left(\phi^{\theta}_{ijl} \tilde{D}_{ijl}^{\theta}(n\delta t)  + \phi^{\gamma}_{ijl} \tilde{D}_{ijl}^{\gamma}(n\delta t)  \right).
 \end{equation}
Here, the effective mass $m_i$, velocity-velocity correlation function $C_i^{vv}(n\delta t)$, the impulse-velocity correlation function $C_i^{Iv}(n\delta t)$, and the dynamical diffusivity $D_i(n\delta t)$ are computed directly from the trajectory and stored. $\tilde{D}_{ijl}^{\theta}(n\delta t)$ is then calculated with the numerical integration $\tilde{D}_{ijl}^{\theta}(n\delta t)=\delta t \sum\limits_{s=0}^{n-1} \theta_{jl}((n-s-\frac{1}{2})\delta t)D_i((s+\frac{1}{2})\delta t)$ and stored. The same procedure also applies to $\tilde{D}_{ijl}^{\gamma}(n\delta t)$. 
After these preparations, it is straightforward to optimize the loss function with the gradients:
\begin{equation}
    \frac{\partial \mathcal{L}_Y[\tau, \sigma^{\theta/\gamma}]}{\partial \tau} =  -\frac{\delta t}{T_{\mathrm{cut}}}   \sum_{n=1}^N  \sum_{ijl} 2m_i\varepsilon_i(n\delta t)   \left(\frac{\partial\phi^{\theta}_{ijl}}{\partial \tau} \tilde{D}_{ijl}^{\theta}(n\delta t)  + \frac{\partial \phi^{\gamma}_{ijl}}{\partial \tau} \tilde{D}_{ijl}^{\gamma}(n\delta t)  \right)
\end{equation}
and 
\begin{equation}
    \frac{\partial \mathcal{L}_Y[\tau, \sigma^{\theta/\gamma}]}{\partial \sigma^{\theta/\gamma}} =  -\frac{\delta t}{T_{\mathrm{cut}}}  \sum_{n=1}^N \sum_{ijl}  2m_i\varepsilon_i(n\delta t)  \left(\frac{\partial\phi^{\theta}_{ijl}}{\partial \sigma^{\theta/\gamma}} \tilde{D}_{ijl}^{\theta}(n\delta t)  + \frac{\partial \phi^{\gamma}_{ijl}}{\partial \sigma^{\theta/\gamma}} \tilde{D}_{ijl}^{\gamma}(n\delta t)  \right).
\end{equation}
For practical implementation, $\frac{\partial\phi^{\theta}_{ijl}}{\partial \tau}$, $\frac{\partial\phi^{\theta}_{ijl}}{\partial \sigma^{\theta}}$ and $\frac{\partial\phi^{\theta}_{ijl}}{\partial \sigma^{\gamma}}$ are traced by autodifferention engines. Their exact expressions are omitted here. 

\subsection{The friction constant in ab initio Langevin equation}
For a given AIGLE with the memory kernel $K_i(s) = \sum_{jl} \phi^{\theta}_{ijl} \theta_{jl}(s) + \phi^{\gamma}_{ijl} \gamma_{jl}(s)$, the friction constant in the corresponding AILE is given by 
\begin{equation}
 \eta_i = -\int_0^\infty K_i(s) ds  = -\sum_{jl} \left(\phi^{\theta}_{ijl} \int_0^\infty \theta_{jl}(s)ds + \phi^{\gamma}_{ijl} \int_0^\infty\gamma_{jl}(s) ds\right). 
\end{equation}
With $\lambda_j=1/\tau_j$ and $\alpha_{jl}=\tfrac{2\pi (l-1)}{L\tau_j}$, the closed-form expression of $\eta_i$ is given as
\begin{equation}
    \eta_i = -\sum_{j=1}^J \sum_{l=1}^L \left( \frac{\lambda_j\phi^{\theta}_{ijl}}{\lambda_j^2 + \cos^2 \alpha_{jl}}  +  \frac{\alpha_{jl}\phi^{\gamma}_{ijl}}{\lambda_j^2 + \cos^2 \alpha_{jl}} \right).
\end{equation}

\subsection{Details of training}
{\bf Toy polymer}
The AIGLE model for the toy polymer model is trained on a $10$ns-long all-atom trajectory of the backbone atoms. The size of the decay-Fourier basis is specified by $I=3$, $J=4$. The cutoff is chosen as $T_{\text{cut}}=6$ps.  
The initial condition of $\tau$ is $\tau=(0.1, 0.6, 3)$ps. For training, the optimizer is an ADAM optimizer~\cite{kingma2014adam}. The learning rate is 0.01 for $\tau$  and 0.4 for $\sigma^{\theta/\gamma}$. The learning rates decay by 0.98 for every 100 epochs. The total number of training epochs is 9000. 
At the end, $\tau$ converges to $(0.37, 0.92, 1.26)$ps. A reproducible example of this training is publicly available at Ref.~\cite{aigletools}.

{\bf Alanine dipeptide}
The AIGLE model for dihedral angles is trained on a $60$ns-long all-atom trajectory. The size of the decay-Fourier basis is specified by $I=5$, $J=4$. The cutoff is chosen as $T_{\text{cut}}=100$ps.  
The initial condition of $\tau$ is $\tau=(0.1, 1, 2, 10, 50)$ps. With ADAM optimizer~\cite{kingma2014adam}, the learning rate is 0.01 for $\tau$  and 0.4 for $\sigma^{\theta/\gamma}$. The learning rates decay by 0.98 for every 100 epochs. The total number of training epochs is 9000. 
At the end, $\tau$ converges to $(0.06, 0.12, 2.37, 11.04, 33.04)$ps. A reproducible example of this training is also publicly available at Ref.~\cite{aigletools}.

\end{document}